\documentclass[preprint]{ptephy_v1}

\preprintnumber{XXXX-XXXX} 
\usepackage{hyperref}

\usepackage{amsmath} 
\usepackage{amsthm} 
\usepackage{hyperref} 
\usepackage{graphics} 

\begin{document}

\title{Neutron correlations and clustering in neutron-rich nuclear systems}
\author{Zaihong Yang}
\affil[1]{School of Physics and State Key Laboratory of Nuclear Physics and Technology, Peking University, 
Beijing 100871, China \email{zaihong.yang@pku.edu.cn}}

\author{Yuki Kubota}
\affil[2]{RIKEN Nishina Center for Accelerator-Based Science, 2-1 Hirosawa, Wako 351-0198, Saitama, Japan \email{kubota@ribf.riken.jp}}

\begin{abstract}%
In this paper, we will briefly review the recent progress on neutron correlations and clustering in neutron-rich nuclei from quasi-free scattering experiments. The quasi-free ($p$, $pn$) reaction was measured for Borromean nuclei $^{11}\mathrm{Li}$, $^{14}\mathrm{Be}$, and $^{17}\mathrm{B}$. A surprisingly small $s$-wave component was found for $^{17}$B, revealing a weak neutron halo in $^{17}$B, and the comparative study of the three nuclei shows the surface localization of dineutron correlation and its universality in Borromean nuclei. The two-neutron emission of $^{16}$Be was also studied, finding strong dineutron correlation in $^{16}$Be(g.s.) but weak correlation in $^{16}$Be($2^+$). Two missing-mass measurements using  $^{4}\mathrm{He}$($^{8}\mathrm{He}$, $^{8}\mathrm{Be}$) and $^{8}\mathrm{He}$($p$, $p\alpha$) reactions provided evidence for the $^{4}n$ resonance, 
but the $^{3}n$ resonance was not supported by the recent experiment employing the $t(t,^3\mathrm{He})^3n$ charge-exchange reaction. The quasi-free $(p,p \alpha)$ reaction has been extended to unstable nuclei, and the recent experiment unravels the $\alpha$-$2n$-$\alpha$ molecule-like cluster structure of $^{10}$Be(g.s.).
\end{abstract}

\subjectindex{cluster, neutron correlation, multi-neutron system, knockout}

\maketitle

\section{Introduction} 
The nucleus is generally considered as a collection of neutrons and protons, each moving in a mean field created by the other nucleons.
The fermionic nature of the nucleons and the residual interactions among them further give rise to the well-known shell structure which is similar to that of the electron in atoms. In recent years, extensive efforts have been devoted to exploring and understanding correlations (including also formation of clusters like dineutron and $\alpha$ clusters) beyond the mean-field or shell-model picture. This is not only important for improving our theories about nuclear interactions and structures and reactions of exotic nuclei far off the stability line, but also provides essential information for modeling the macroscopic nuclear systems like neutron stars \citep{Ye2025NPR,Wei2024,oertel2017equations}. Important progress has been achieved, taking advantage of the high-intensity radioactive ion beam facilities such as the Radioactive Isotope Beam Factory (RIBF) of RIKEN, the development of structure and reaction theories, the application of state-of-the-art detector devices.

Formation of clusters, strongly correlated subunits such as $\alpha$ clusters, out of the mean field is a topic of great interest and fundamental importance throughout the history of nuclear physics. In the past years, experimental and theoretical efforts have been mainly focused on the novel cluster states in the excited states of light nuclei---e.g., the $\alpha$-condensate states \citep{Zhou2023nc,Adachi2021plb,Chen2023SB}, the molecular resonant states in beryllium isotopes\citep{Yang2014PRL,Freer2006} and the 3-$\alpha$-linear-chain states in carbon isotopes\citep{HanJX-2023,Yamaguchi2017,LiuYang_2020}. While in light stable nuclei with approximately equal numbers of neutrons and protons formation of cluster states in proximity to the corresponding cluster-separation threshold may be attributed to the saturation property of the nuclear matter, it has remained elusive how the excess neutrons impact clustering when entering the neutron-rich region, and because of the large proton-neutron asymmetry the symmetry energy should play a prominent role here \citep{KIMURA2016}. According to recent theoretical calculations based on the generalized relativistic mean-field model (gRDF), $\alpha$ clusters can also be formed in low-density nuclear matter including the low-density surface region of neutron-rich heavy nuclei, and, more importantly, the cluster formation will lead to inhomogeneity and thus affect the properties of the nuclear matter\citep{TYPEL2014,Typel-2010,OERTEL2017}. Such clustering effect has basically been overlooked previously when constraining the nuclear equation of state. Recent $(p,p \alpha)$ experiments on neutron-rich stable tin isotopes \citep{TANAKA2021} and on the neutron-rich unstable nucleus $^{10}\mathrm{Be}$ \citep{Li-ppa-2023} make an important step towards understanding clustering in neutron-rich nuclei.

Significant progress has been made on the two-neutron correlations and the dineutron cluster ($^{2}n$) in the past decades \citep{nakamura2006observation,kubota2020surface,corsi2023searching}, but only a few experiments on the trineutron cluster ($^{3}n$) and tetraneutron cluster ($^{4}n$) have been reported\citep{marques2002detection,duer2022observation,MikiPRL2024} and contrasting predictions are provided from the state-of-the-art nuclear structure theories\citep{pieper2003can, timofeyuk2003multineutrons, lazauskas2005physically, shirokov2018tetraneutron, fossez2017can, deltuva2018tetraneutron, higgins2020nonresonant, lazauskas2022low}. It has remained an open yet intriguing question as to whether a neutral cluster made purely of neutrons exists, despite extensive experimental and theoretical efforts for more than half a century. The properties of these multi-neutron systems provides a stringent test for the underlying nuclear force, particularly for the isospin-dependent component, and they also provide unique access to neutron-neutron and multi-neutron correlations. 
In experiment, the charge-exchange reaction, multinucleon-transfer reaction, and nucleon/cluster knockout reactions such as ($p$, $2p$) ($p$, $p\alpha$) ($p$, $3p$) are currently utilized for the production of such exotic systems (see recent review article \citep{marques2021quest} for more details). So far, these experiments are all based on the missing-mass method without detection of the decay neutrons, which does not allow to investigate the inner correlations and structure of these multi-neutron clusters.

\section{Halos and neutron correlations in light neutron-rich nuclei}
For neutron-rich nuclei around the neutron drip-line region, the valence neutrons---when occupying the $s$ or $p$ orbital---can tunnel far out into the ``classically-forbidden" region and thus give rise to the neutron halo. Since the first discovery by Tanihata $et$ $al.$ in the 1980s \citep{tanihata1985measurements, Hansen1987}, the neutron halo has been at the focus of both experimental and theoretical studies (see, $e.g.$, recent review articles \citep{TANIHATA2013,FREDERICO2012}). Of particular interest are two-neutron halo nuclei (also known as Borromean nuclei) such as $^{11}$Li, for which the correlation between the two valence neutrons is crucial for the binding. They will help to understand the neutron correlations and their density dependence in neutron-rich matter which hitherto basically depends on the predictions of mean-field theories. 

\begin{figure}[t!]
\setlength{\abovecaptionskip}{0pt}
\setlength{\belowcaptionskip}{0pt}
\begin{center}
\includegraphics[trim = 0 0 0 0,clip,angle=0,width=6.0in]{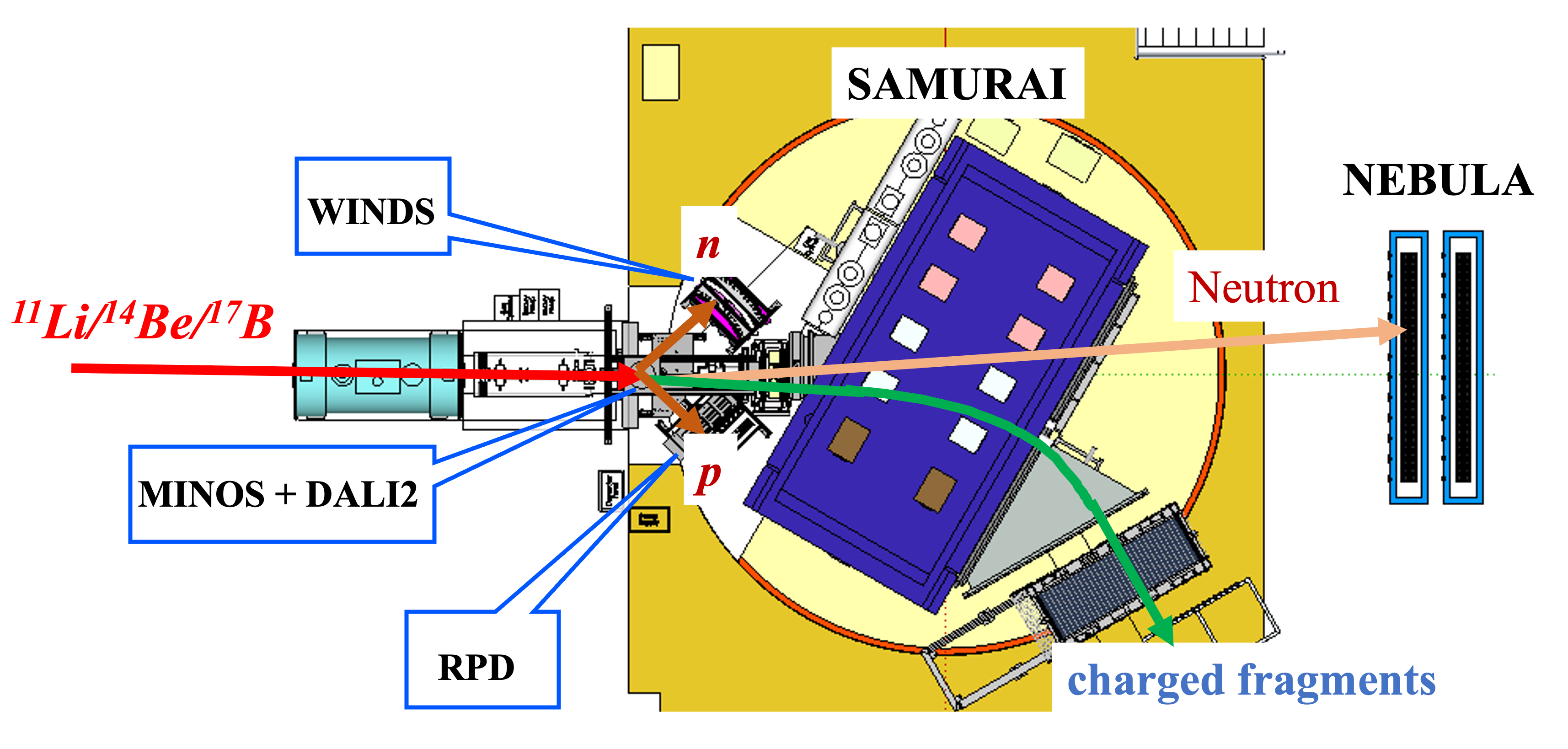}
\end{center}
\caption{Schematic view of the experimental setup for the $(p,pn)$ experiment at RIBF.}
\label{setup}
\end{figure}

A series of $(p,pn)$ experiments were carried out at RIBF to study the halo structure, the neutron-neutron correlation and formation of dineutron clusters under strong correlations \citep{kubota2020surface,Yang2021prl,corsi2023searching}. These studies concern a kinematically complete measurement, which was made possible by combining the high-intensity beams neutron-rich unstable nuclei provided by RIBF and the state-of-the-art detector devices including the vertex-tracking liquid hydrogen target MINOS \citep{Obertelli2014}, in-beam $\gamma$-ray spectrometer DALI2 \citep{TAKEUCHI2014}, and the SAMURAI spectrometer \citep{KOBAYASHI2013,KONDO2020}. A schematic view of the experimental setup is presented in Fig. \ref{setup}. After the $(p,pn)$ reaction on MINOS, the recoil proton was analyzed by the TPC of MINOS and the recoil proton spectrometer, and the recoil neutron partner was detected by the neutron detector array WINDS. The charged fragments and decay neutrons were analyzed by the SAMURAI spectrometer \citep{KOBAYASHI2013}and the neutron detector array NEBULA \citep{KONDO2020}. The DALI2 in-beam $\gamma$-ray spectrometer \citep{TAKEUCHI2014} was also installed at the target region for $\gamma$-ray detection. 

\subsection{Halo structure of  $^{17}\mathrm{B}$ }
\begin{figure}[t!]
\setlength{\abovecaptionskip}{0pt}
\setlength{\belowcaptionskip}{0pt}
\begin{center}
\includegraphics[trim = 0 0 0 0,clip,angle=0,width=4.0in]{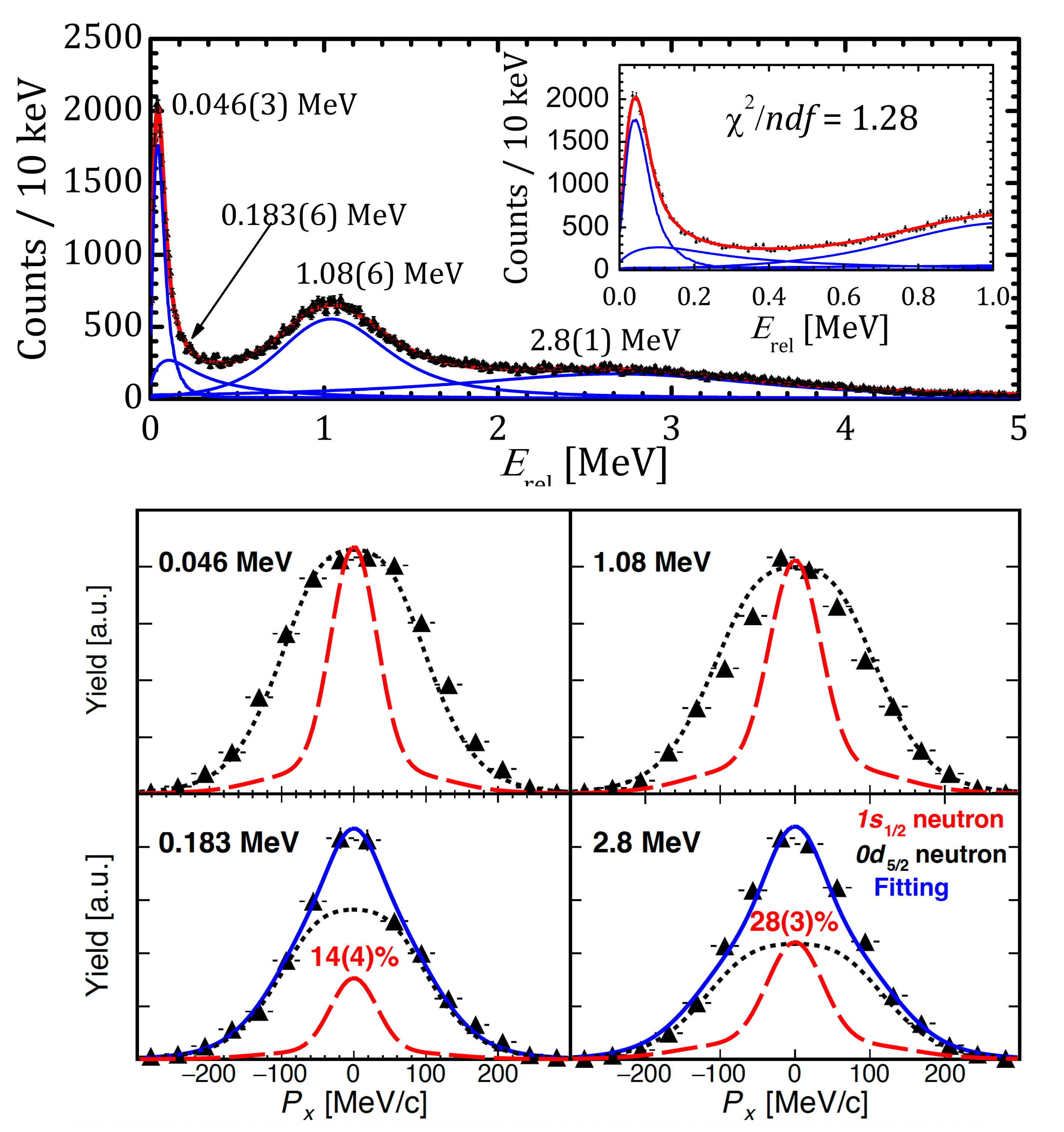}
\end{center}
\caption{(Top panel) The $E_{\mathrm{rel}}$ spectrum populated in the $^{17}\mathrm{B}(p,pn)$ experiment together with the fitting with a sum of four resonances. (Bottom panel) Transverse momentum ($P_\mathrm{x}$) distributions for different $^{16}\mathrm{B}$ states. The figure is adapted from \citep{Yang2021prl}.}
\label{b17result}
\end{figure}

Halo features in $^{17}$B have been well known in prior experiments---the large matter radius \citep{OZAWA2001NPA}, the narrow momentum distribution of $^{15}$B fragments \citep{Suzuki1999prl}, and the thick neutron surface \citep{Estrad2014prl}, but the single-particle configuration of the valence neutrons has not been well 
determined. A large $s$-orbital percentage of $\sim 50 \%$ was deduced from the matter radius or the $^{15}$B core momentum distribution based on analysis using a $^{15}$B+$n$+$n$ 3-body model with $^{15}$B being an inert and spherical core \citep{OZAWA2001NPA,Suzuki1999prl,Fortune2012}. However, Estrad\'{e} {\it et al.} found that such a 3-body picture bearing out a dominant $s$ orbital component was inconsistent with their new results of the neutron skin thickness \citep{Estrad2014prl}. Yang $et$ $al.$ measured the $0d _{5/2}$ and $1s _{1/2}$ spectroscopic factors in $^{17}$B using the quasi-free $^{17}\mathrm{B}$($p$, $pn$) reaction in inverse kinematics \citep{Yang2021prl}. By analyzing the momentum distribution as shown in Fig. \ref{b17result}, the exclusive cross sections for each of the $^{16}$B states populated in $^{17}$B were extracted. The spectroscopic factors for $1s_{1/2}$ and $0d_{5/2}$ orbitals were thus obtained by comparing with the predicted single-particle cross sections of DWIA calculations, unraveling a surprisingly small $1s_{1/2}$ percentage of 9(2)$\%$ in $^{17}$B. The finding of such a small $1s_{1/2}$ component and the halo features reported in prior experiments reveal a definite but weak neutron halo in $^{17}$B. This work gives the smallest percentage of $s$ or $p$ orbitals among known halo nuclei, and implies that the dominant occupation of $s$ or $p$ orbitals is not a prerequisite for the occurrence of neutron halo, as suggested by Hove $et$ $al.$ for very weakly bound systems \citep{Hove2018prl}. 

\subsection{Dineutron universality in Borromean nuclei $^{11}\mathrm{Li}$, $^{14}\mathrm{Be}$, and $^{17}\mathrm{B}$}
\begin{figure}[t!]
\setlength{\abovecaptionskip}{0pt}
\setlength{\belowcaptionskip}{0pt}
\begin{center}
\includegraphics[trim = 0 0 0 0,clip,angle=0,width=5.0in]{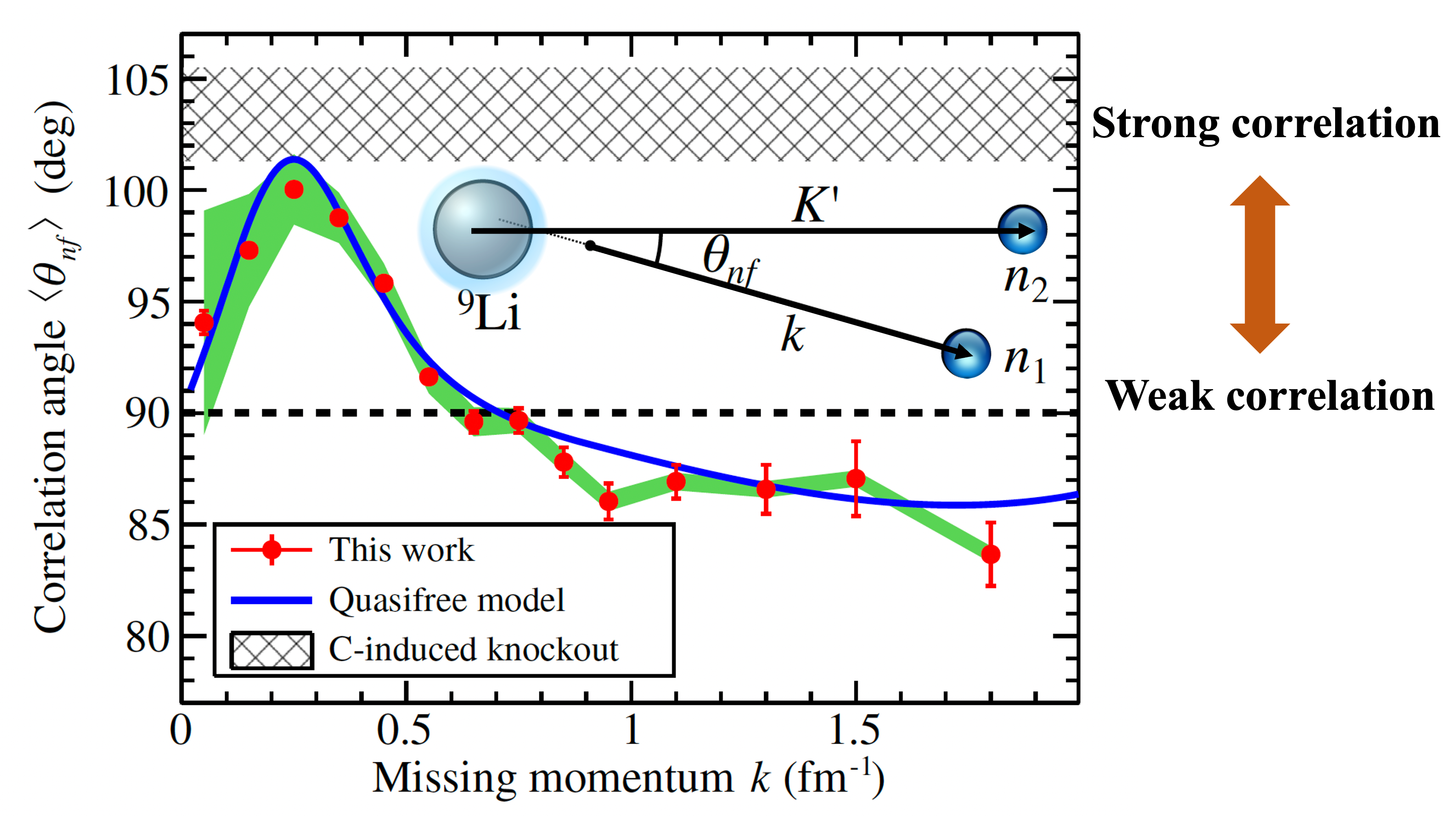}
\end{center}
\caption{Change of the mean value of the correlation angle $\cos \theta_{nf}$ with respect to the missing momentum $k$ of the knocked-out neutron. Red points show the experimental data of \citep{kubota2020surface}. The blue curve shows the quasifree model calculation. Black hatched area shows the result of a previous study using carbon target for comparison \citep{Simonprl1999}. The figure is adapted from \citep{kubota2020surface}.}
\label{li11result}
\end{figure}

Dineutron ($^{2}n$) generally refers to a spatially compact neutron pair with a total spin of 0 \citep{migdal19732}. The neutron halo is characterized by a spatially extended neutron distribution around a tightly bound core. Two-neutron halo nuclei thus serve as an excellent laboratory for studying the $^{2}n$ correlations since the neutron-neutron correlation and consequently the formation of $^{2}n$ are expected to be enhanced in the dilute neutron matter of the neutron halo \citep{matsuo2006spatial}. The most notable example is $^{11}\mathrm{Li}$ with a very small separation energy of $S_{2n}$ = 369 keV and a well-developed halo structure \citep{matsuo2006spatial, smith2008first, tanihata1985measurements}. Strong dineutron correlation in $^{11}\mathrm{Li}$ was predicted by both the three-body model and the Gamow couple-channel model calculations \citep{hagino2007coexistence, Chen2021prc}. Experimentally, it was probed by Nakamura $et$ $al.$ by measuring the electric dipole ($E1$) response in the Coulomb dissociation experiment. From the measured $B(E1)$ strength, an opening angle of $\sim48$ degrees was determined for the two valence neutrons with respect to the core, indicating strong dineutron correlations in the ground state of $^{11}\mathrm{Li}$ \citep{nakamura2006observation}. However, the $n-n$ final state interaction may also play an important role in the Coulomb dissociation of two-neutron halo nuclei \citep{Kikuchi2010}. To minimize the FSI effect, Kubota $et$ $al.$ carried out a kinematically complete measurement of the $^{11}\mathrm{Li}$($p$, $pn$) reaction under the quasi-free condition \citep{kubota2020surface}. Taking advantage of the high statistics and the high resolution of the detector setup, the two-neutron configurations in $^{11}\mathrm{Li}$ were determined as $35(2)\% s^2 +59(1)\% p^2+6(4)\% d^2$ by analyzing the missing momentum of the knocked-out neutron for each $^{9}\mathrm{Li}+n$ relative energy bin. More importantly, the use of proton target allows to probe the neutron-neutron correlations in a large volume of $^{11}\mathrm{Li}$---from the low-density halo region to the internal region around the saturation density. To facilitate direct comparison with the predictions of the quasi-free reaction model based on a $^{9}\mathrm{Li}+n+n$ 3-body structure of $^{11}\mathrm{Li}$, the correlation angle $\theta_{nf}$ was adopted as a measure of the neutron-neutron correlation. As shown in Fig.\ref{li11result}, Kubota $et$ $al.$ analyzed the evolution of the mean value of $ \cos \theta_{nf}$ with respect to the missing momentum $k$ of the knocked-out neutron which is inversely correlated with the density. The result revealed an interesting density-dependent behavior of the $n-n$ correlation---namely, the dineutron correlation is enhanced in a limited low-density region around the $^{11}\mathrm{Li}$ surface but gets suppressed when entering the inner region with higher density or the more dilute halo region with further lower density. This density-dependent behavior of $^{2}n$ formation and $n-n$ correlations in general is consistent with the predictions of Hartree-Fock-Bogoliubov calculaitons by Matsuo for infinite nuclear matter \citep{matsuo2006spatial}. 

This finding was further corroborated by the comparative study of dineutron correlation in the three typical Borromean nuclei $^{11}\mathrm{Li}$, $^{14}\mathrm{Be}$, and $^{17}\mathrm{B}$ that exhibit different degrees of halo structure by Corsi $et$ $al.$ \citep{corsi2023searching}. In this work, the average correlation angle as a function of the intrinsic momentum of the removed neutron was analyzed to probe the location and evolution of the dineutron correlation, similar to \citep{kubota2020surface}, and the result reveals the dineutron correlation in the peripheral region of $^{14}\mathrm{Be}$ and $^{17}\mathrm{B}$, but it gets damped compared to $^{11}\mathrm{Li}$. This work also indicates the universality of dineutron correlation in the low-density surface of Borromean nuclei. It is thus very interesting to extend such ($p$, $pn$) experiments to heavier neutron-rich nuclei, and to understand the surface localization or density-dependence of dineutron correlation it is also very helpful to combine measurements using complementary reaction probes. The damping of dineutron correlation in $^{14}\mathrm{Be}$ could be due to the presence of core-excited configurations, according to the three-body model calculations \citep{corsi2023searching}. Actually, the core-excited configuration of  $^{11}\mathrm{Li}$ has also been measured in the $^{11}\mathrm{Li}$($p$, $pn$) experiment of Kubota $et$ $al.$ \citep{kubota2020surface}. The data analysis is still ongoing, aiming to clarify the effect of core-excited configurations on the dineutron correlation in  $^{11}\mathrm{Li}$.

\subsection{Two-neutron decay of $^{16}\mathrm{Be}$ }
\begin{figure}[t!]
\setlength{\abovecaptionskip}{0pt}
\setlength{\belowcaptionskip}{0pt}
\begin{center}
\includegraphics[trim = 0 0 0 0,clip,angle=0,width=4.0in]{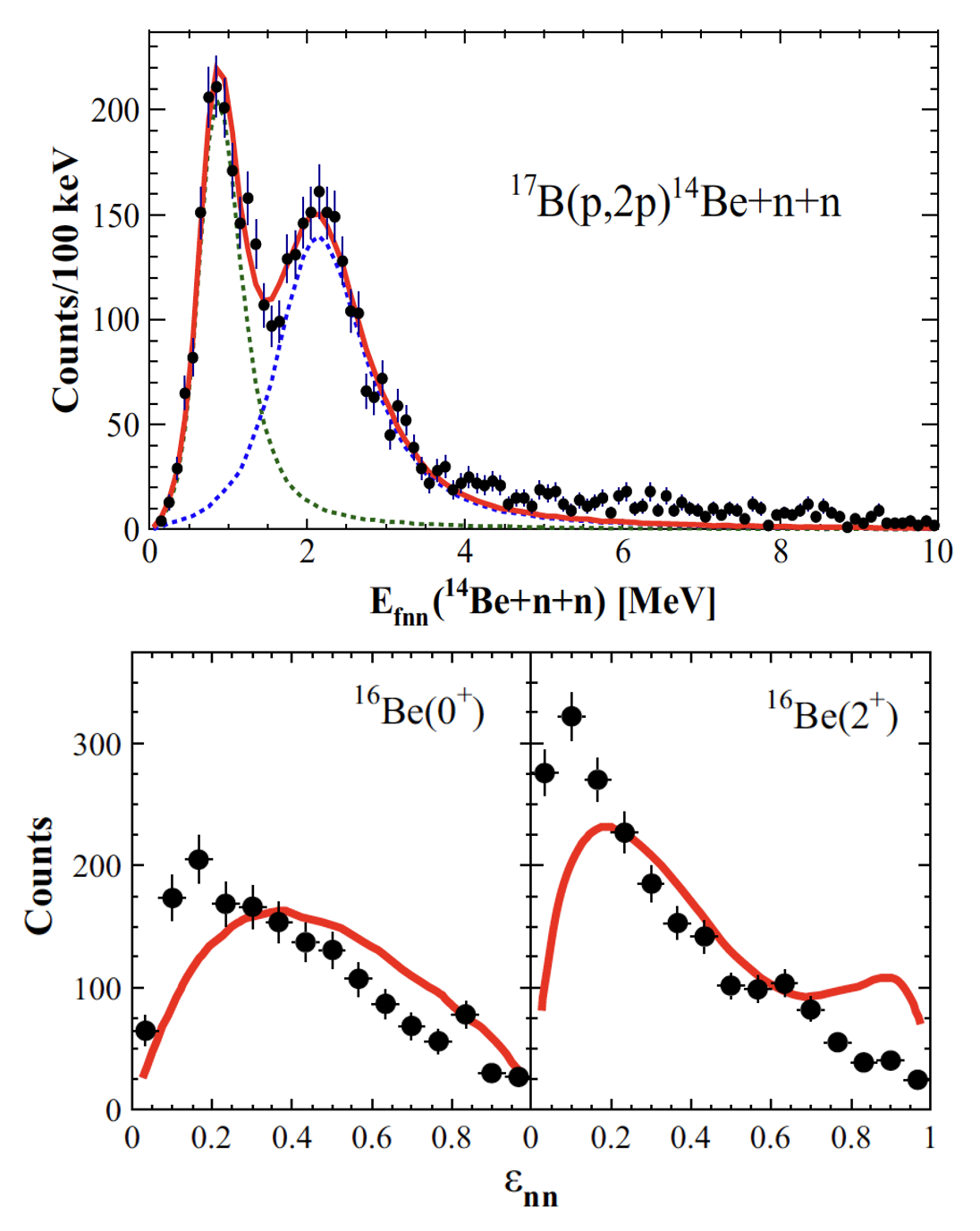}
\end{center}
\caption{(Top panel) The $E_{\mathrm{rel}}$ spectrum of $^{16}\mathrm{Be}$ populated in the $^{17}\mathrm{B}(p,2p)$ experiment together with fitting with a sum of two resonant states. (Bottom panel) The normalized $n-n$ relative energy distribution for the $0^+$ and $2^+$ states of $^{16}\mathrm{Be}$, in comparison to the prediction of the three-body model calculations. The figure is adapted from \citep{Monteagudo2024prl}.}
\label{Be16result}
\end{figure}

Two-neutron decay has been measured for unbound nuclei, such as $^{13}\mathrm{Li}$ \citep{johansson2010three, kohley2013first} and $^{16}\mathrm{Be}$ \citep{spyrou2012first}, and the excited states of neutron-rich nuclei \citep{marques2001three, aksyutina2013study}. Among them, $^{16}\mathrm{Be}$ has garnered great interest for search of direct dineutron emission since the sequential two-neutron emission via $^{15}\mathrm{Be}$ is energetically suppressed. 
By comparing the observed $n$-$n$ energy and angular correlation patterns with the model calculations assuming different two-neutron emission processes, Spyrou $et$ $al.$ found evidence for direct dineutron emission of the ground state of $^{16}\rm{Be}$ \citep{spyrou2012first}. 
However, possible effects of the $n$-$n$ FSI in the two-neutron decay has been udner hot debate; in \citep{marques2012comment}, Marqu{\'e}s $et$ $al.$ argued that the enhancement at low $n-n$ relative energies in \citep{spyrou2012first} could also be explained by the direct three-body breakup model incorporating the $n$-$n$ FSI, as an alternative to the dineutron emission model of \citep{spyrou2012first}. 
New data for the structure and $n-n$ correlation of $^{16}\rm{Be}$ were obtained from the $(p,2p)$ reaction on $^{17}\rm{B}$ \citep{Monteagudo2024prl}. The high statistics and high resolution of this new measurement allowed the unambiguous identification of two narrow resonant states of $^{16}\rm{Be}$---the $0^+$ ground state and the $2^+$ first excited state---which were not separated in \citep{spyrou2012first}. To explore the two-neutron decay process and its connection to the initial dineutron cluster structure, a three-body model describing both the initial structure of $^{16}\rm{Be}$ and its time evolution was employed. The three-body model predicts a dominant dineutron cluster structure in the $0^+$ state but weak $n-n$ correlations in $2^+$. The normalized $n-n$ relative energy distributions predicted by the three-body model were then compared to the experimental data (Fig. \ref{Be16result}). For both states of $^{16}\rm{Be}$, the principle patterns of the experimental data were fairly reproduced, and the dineutron cluster in $^{16}\rm{Be}$(g.s.) was thus evidenced. We also mention that the two-neutron decay of $^{11}\rm{Li}$ and $^{13}\rm{Li}$ have been studies recently by using the same experimental setup and theoretical tools \citep{ANDRE2024plb}. In general, the experimentally observed correlation pattern should be determined by both the initial structure and the decay process (including the FSI effects) as discussed in \citep{wang2021fermion}, but it has still remained a challenge for theories to separate the FSI effect from the initial $^{2}n$ structure. It is thus very important to employ microscopic theories that can provide realistic wave function for the initial state and properly describe its time evolution to describe the two-neutron decay. Important progress has been achieved in recent years, such as the time-dependent method based on the Gamow coupled-channel model \citep{wang2021fermion} and the time-evolution three-body model used in the study of $^{16}\rm{Be}$ \citep{Monteagudo2024prl}, and new data of $n-n$ correlation with improved detector resolutions are needed to benchmark the theoretical models. 

\section{Multineutron systems}
Explorations on heavier neutron clusters ($^{3}n$, $^{4}n$ ...) beyond $^{2}n$ are almost at the limits of present experimental capabilities due to the limited radioactive beam intensities and extremely low multi-neutron detection efficiency. In the early search for $^{3}n$ and $^{4}n$ using double-charge-exchange reactions ($\pi^{-}$, $\pi^{+}$) \citep{ungar1984search, gorringe1989search, grater1999search} and multi-nucleon-transfer reactions \citep{ohlsen1968search, cerny19747li+, belozyorov1988search, bohlen1995study}, significant effects of neutron correlations in the multi-neutron systems could be inferred from the observed missing-mass spectrum through comparison to the phase-space continuum spectrum, but these experiments fall short of being conclusive on the presence of neutron cluster states. Below we will discuss three latest experiments for $^{3}n$ and $^{4}n$ based on the high-intensity radioactive ion beams of RIBF.
\subsection{Tetraneutron ($^{4}n$)}
At the beginning of the new century, Marqu{\'e}s $et$ $al.$ performed a breakup experiment of the neutron-rich unstable nucleus $^{14}\rm{Be}$ to search for $^{4}n$ at GANIL\citep{marques2002detection, marques2005possible}. Several events recorded by the neutron detector array were found consistent with a bound or resonant $^{4}n$ state. Soon after this experiment, the existence of a bound $^{4}n$ was ruled out by ab initio Green’s function Monte Carlo (GFMC) calculations using realistic nuclear forces \citep{pieper2003can}. But no consensus was reached about the existence of a $^{4}n$ resonant state \citep{pieper2003can, timofeyuk2003multineutrons, lazauskas2005physically}. The above mentioned GFMC calculation of S. C. Pieper predicted a broad $^{4}n$ resonance at around 2 MeV \citep{pieper2003can}. On the contrary, the few-body calculation of Lazauskas $et$ $al.$ based on Faddeev-Yakubovsky (FY) equations in conﬁguration space using the modern nuclear Hamiltonians found that tetraneutron resonances could hardly exist \citep{lazauskas2005physically}.

In 2016, new experimental evidence for the $^{4}n$ resonance was reported in a double-charge-exchange reaction $^{4}\mathrm{He}$($^{8}\mathrm{He}$, $^{8}\mathrm{Be}$) carried out by Kisamori $et$ $al.$ at RIBF \citep{kisamori2016candidate}. The SHARAQ spectrometer was utilized to conincidently measure the two $\alpha$ particles from the decay of $^{8}\mathrm{Be}$, and candidate $^{4}n$ events were selected by gating on the invariant-mass spectrum of the two $\alpha$ particles to select the ground state of $^{8}\mathrm{Be}$. The mass of $^{4}n$ was then deduced by using the missing-mass method based on the momenta of the $^{8}\mathrm{He}$ projectile and the recoil $^{8}\mathrm{Be}$. The measurement was focused on the backward angles to populate the $^{4}n$ state under the recoil-less condition. The background is largely reduced taking advantage of the high resolution of the SHARAQ spectrometer and the high selectivity of the reaction channel of interest. After careful analysis of the continuum background, four events right above the breakup threshold on the missing-mass spectrum were interpreted as a candidate resonant $^{4}n$ state with a significance level of 4.9$\sigma$. The resonant energy was obtained as 0.83 $\pm$ 0.65 (stat) $\pm$ 1.25 (syst) MeV and an upper limit of 2.6 MeV (FWHM) was estimated for the width. 
\begin{figure}[t!]
\setlength{\abovecaptionskip}{0pt}
\setlength{\belowcaptionskip}{0pt}
\begin{center}
\includegraphics[trim = 0 0 0 0,clip,angle=0,width=6.0in]{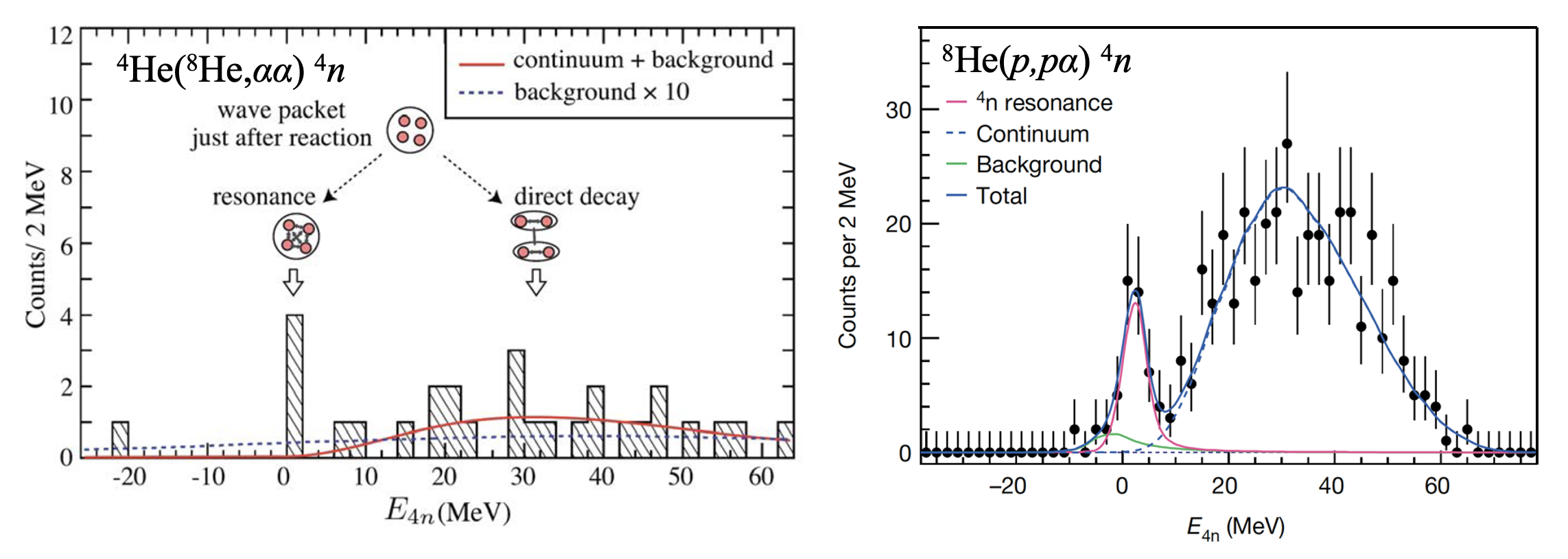}
\end{center}
\caption{The missing-mass spectra of the tetraneutron system obtained from the double-charge-exchange reaction $^{4}\mathrm{He}$($^{8}\mathrm{He}$, $^{8}\mathrm{Be}$) (left panel) \citep{kisamori2016candidate} and from the knockout reaction $^{8}\mathrm{He}$($p$, $p\alpha$) (right panel).
\citep{duer2022observation}. The figure is adapted from \citep{kisamori2016candidate,duer2022observation}.}
\label{4nresult}
\end{figure}

This intriguing experimental result spurred renewed interest in multi-neutron states. The no-core shell model (NCSM) employing realistic two-body interaction JISP16 by Shirokov $et$ $al.$ predicted the $^{4}n$ state with a resonant energy $E_{4n}$ = 0.84 MeV and width $\Gamma$ = 1.38 MeV \citep{shirokov2016prediction}, in good agreement with the result of Kisamori $et$ $al.$ \citep{kisamori2016candidate}. The results of their latest calculation using the modern $NN$ interactions Daejeon16 and chiral N$^{3}$LO was also similar \citep{shirokov2018tetraneutron}. The $\textit{ab initio}$ no-core Gamow shell model (NCGSM) with the density matrix renormalization group method \citep{fossez2017can} predicted a broad resonance-like four-neutron state with a width of $\Gamma$ $\approx$ 3.7 MeV that was much larger than the reported value of Ref. \citep{kisamori2016candidate}. The authors of Ref. \citep{fossez2017can} speculated that the large width of the tetraneutron made it difficult for the direct observation of this state and that the low-energy peak observed by Kisamori $et$ $al.$ could be attributed to a feature of the four-neutron scattering rather than a genuine resonant state. Interestingly, similar conclusions were also obtained by Deltuva in calculations using the Faddeev–Yakubovsky (FY) and Alt–Grassberger–Sandhas (AGS) formalisms \citep{deltuva2018tetraneutron} and by Higgins $et$ $al.$ using the adiabatic hyperspherical method \citep{higgins2020nonresonant}. On the other hand, the latest $\textit{ab initio}$ NCGSM based on the chiral N$^{3}$LO two-body nuclear force by Li $et$ $al.$ predicted a low-lying $^{4}n$ resonance with a resonance energy of 2.64 MeV and a width of 2.38 MeV \citep{li2019ab}. Hiyama $et$ $al.$ made few-body calculations using the Gaussian Expansion Method and found that an additional strongly attractive $T = 3/2$ isospin-dependent three-body force---that is remarkably inconsistent with the known properties of typical light nuclei like $^{4}\mathrm{He}$---was required to generate an observable $^{4}n$ resonant state \citep{hiyama2016possibility}. 


Recently, Duer $et$ $al.$ observed a low-lying $^{4}n$ state in the $^{8}\mathrm{He}$($p$, $p\alpha$) experiment performed at RIBF \citep{duer2022observation}. The well-developed cluster structure of $^{8}\mathrm{He}$ with an $\alpha$ particle plus four valence neutrons provides unique access to the $^{4}n$ system via the removal of the $\alpha$ particle. The detector setup based on the SAMURAI spectrometer was optimized for high momentum transfer to ensure that the $\alpha$ particle was removed from the incident $^{8}\mathrm{He}$ under the quasi-free ($p$, $p\alpha$) condition and the $^{4}n$ system was thus populated in an unperturbed way. The energy of $^{4}n$ was reconstructed using the missing-mass method. A resonance-like peak near the threshold was clearly observed, with a significance level well beyond 5$\sigma$. By fitting the missing-mass spectrum, the resonant energy ($E_{4n}$) and width ($\Gamma$) were obtained, $E_{4n}$ = 2.37 $\pm$ 0.38 (stat) $\pm$ 0.44 (syst) MeV and $\Gamma$ = 1.75 $\pm$ 0.22 (stat) $\pm$ 0.30 (syst) MeV, which are compatible with the results of Kisamori $et$ $al.$ but with significantly higher statistics. 

Soon after the report of Duer $et$ $al.$, Lazauskas $et$ $al.$  proposed an alternative explanation for the observed resoance-like peak in the missing-mass spectrum of the $^{8}\mathrm{He}$($p$, $p\alpha$) reaction \citep{lazauskas2022low}. Lazauskas $et$ $al.$ constructed a reaction model based on AV18 and N$^{3}$LO chiral nuclear forces to describe the $^{8}\mathrm{He}$($p$, $p\alpha$) reaction, and the low-energy peak observed by Duer $et$ $al.$ was well reproduced by their model and was thus attributed to the effect of the initial $\alpha+2n+2n$ cluster structure of $^{8}\mathrm{He}$ and the reaction mechanism ($e.g.$, the final-state interaction among the four neutrons) rather than the formation of a $^{4}n$  resonance. 

We also mention that a hint of $^{4}n$ resonant state was reported recently by Faestermann $et$ $al.$ using the multi-nucleon-transfer reaction $^{7}\mathrm{Li}$($^{7}\mathrm{Li}$, $^{10}\mathrm{C}$) \citep{faestermann2022indications}. 

\subsection{Trineutron ($^{3}n$)}
The trineutron resonance was predicted by the $\textit{ab initio}$ NCGSM and GFMC calculations\citep{li2019ab,gandolfi2017trineutron}---both predicting a $^3n$ resonance even lower than $^4n$. Recently, Mazur $et$ $al.$ employed the SS-HORSE–NCSM method with Daejeon16 and N3LO chiral nuclear interactions to study the possible formation of $^3n$ resonances, predicting two overlapping resonances with $E_{3n} \sim$ 0.5 MeV and $\Gamma \sim$ 1 MeV \citep{Mazur2024}. The experimental search for the trineutron has been challenging. One of the most notable efforts comes from the charge-exchange reaction $t(t,^3\mathrm{He})^3n$ recently performed at RIBF by Miki $et$ $al.$ \citep{MikiPRL2024}. To increase the luminosity, the authors developed a thick Ti-tritium target and used a very high intensity of triton beam from the BigRIPS beam line. The measurement was focused on the low-momentum-transfer region down to $\sim$ 15 MeV/c, which is beneficial for producing fragile systems like $^3n$. The energy spectrum of $^3n$ was constructed using the momentum of the exit $^3\mathrm{He}$ measured by the SHARAQ spectrometer. In contrast to the $^4n$ experiments, no peak structure was observed for the $^3n$ system in the energy spectrum up to $\sim$ 20 MeV above the three-neutron threshold. A comparative study of the $^3p$ system was also conducted by measuring the $^3\mathrm{He}(^3\mathrm{He},t)^3p$ reaction at Research Center for Nuclear Physics (RCNP), Osaka University. Similarly, no low-lying peak structure was observed in the energy spectrum, indicating mostly the nonresonant feature of the $^3p$. Interestingly, significant correlations among the three neutrons of the $^3n$ system---albeil not strong enough to produce a $^3n$ resonance---were evidenced by comparing the data to theoretical calculations assuming different levels of correlations.
\begin{figure}[t!]
\setlength{\abovecaptionskip}{0pt}
\setlength{\belowcaptionskip}{0pt}
\begin{center}
\includegraphics[trim = 0 0 0 0,clip,angle=0,width=4.0in]{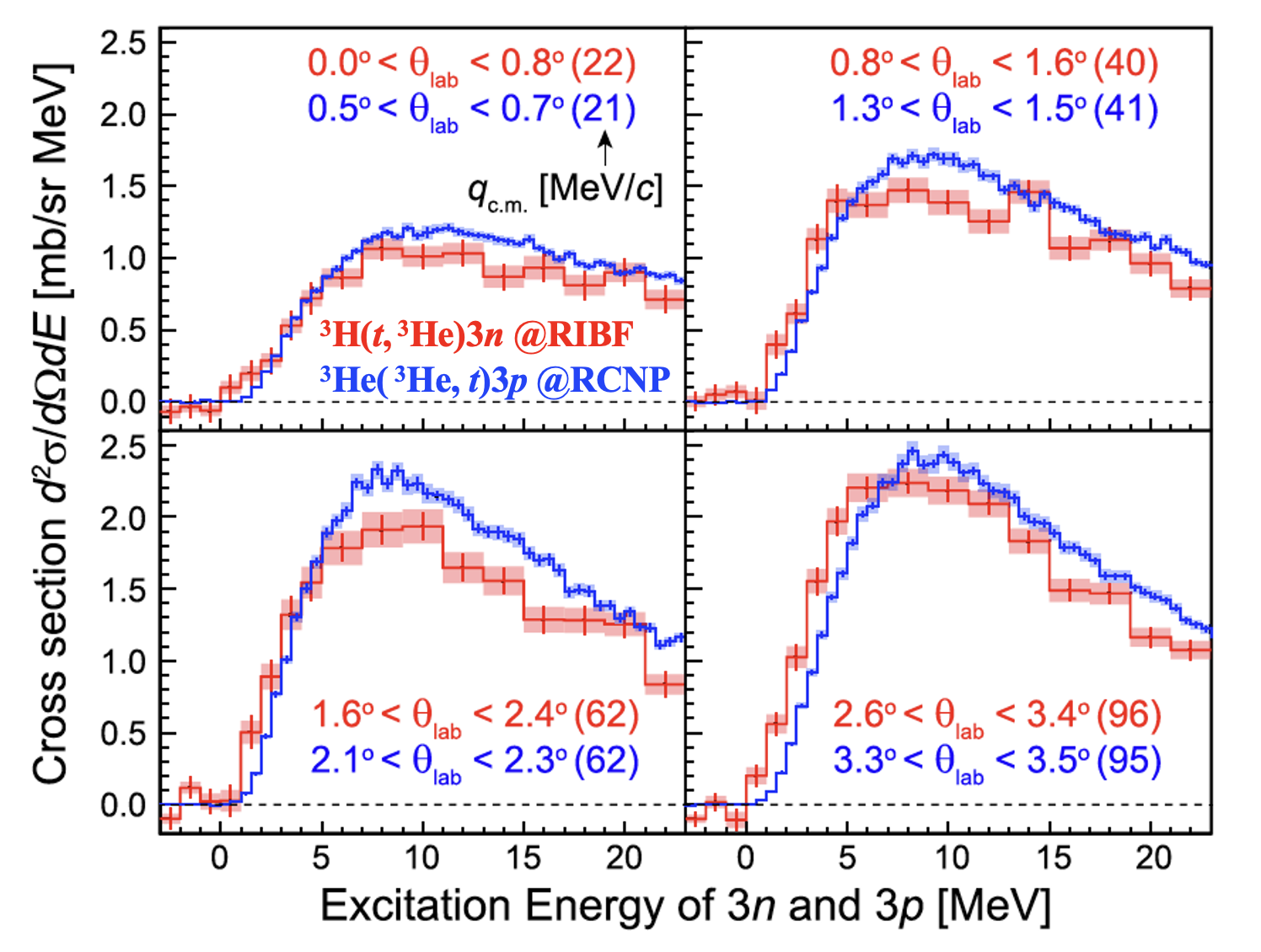}
\end{center}
\caption{The missing-mass spectra of the $3n$ system obtained from the $t(t,^3\mathrm{He})^3n$ reaction at RIBF(red) and that of the $3p$ obtained from the $^3\mathrm{He}(^3\mathrm{He},t)^3p$ reaction at RCNP (blue) at different momentum transfers $q_{c.m.}$ \citep{MikiPRL2024}. The figure is adapted from \citep{MikiPRL2024}.}
\label{4nresult}
\end{figure}

\section{Clustering in nuclear ground states probed using knockout reactions}  
\begin{figure}[!http]
\centering
\includegraphics[trim = 0 0 0 0,clip,angle=0,width=5.0in]{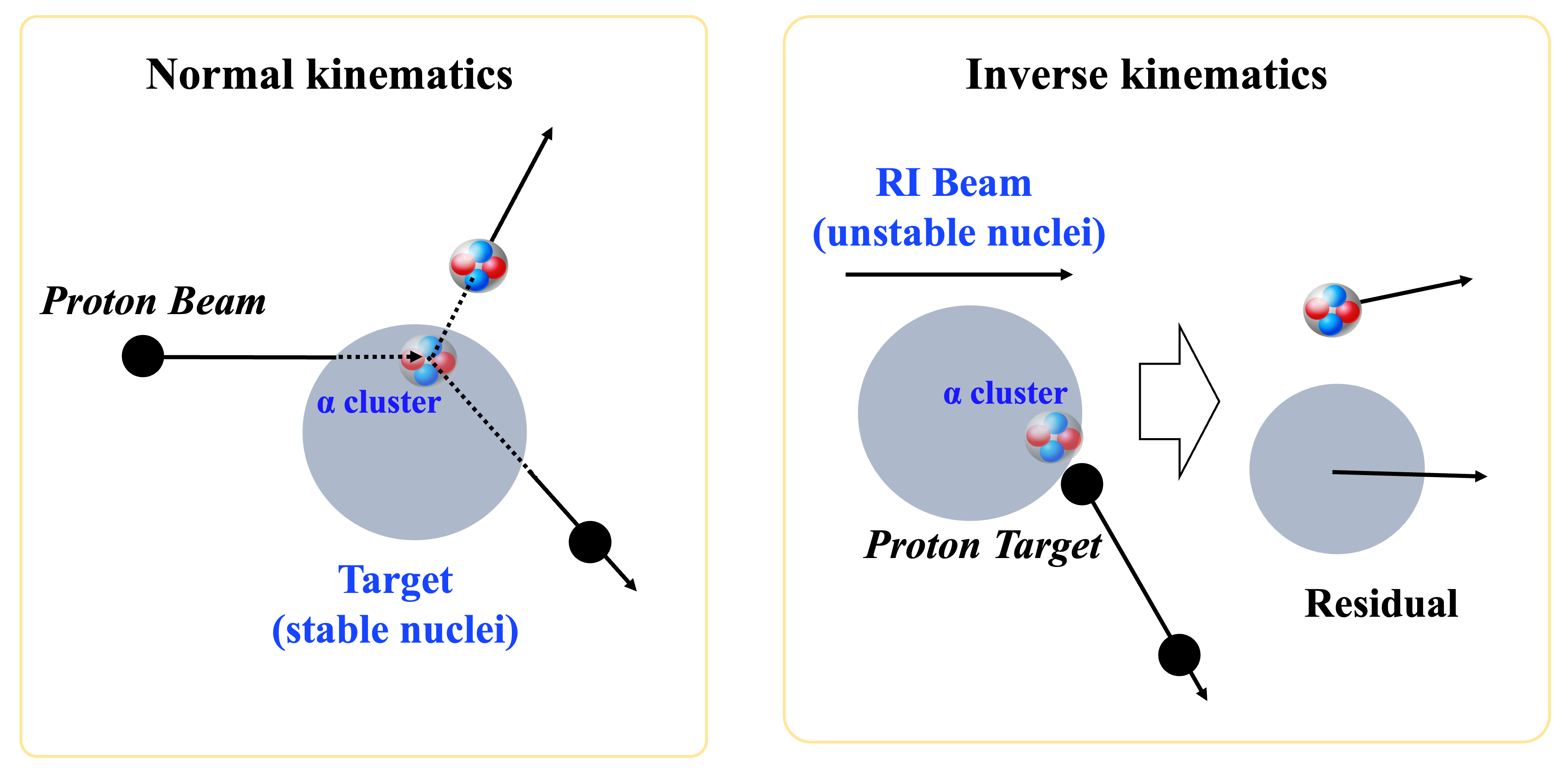}
\caption{\footnotesize{Schematic view of the quasi-free $(p,p\alpha)$ reaction in normal kinematics (left panel) and inverse kinematics (right panel).}}
\label{QFSppa}
\end{figure}

$\alpha$ clusters, when unbound, can be spontaneously emitted taking advantage of the quantum tunneling effect. This has been well known in $\alpha$ decay of heavy radioactive nuclei and the excited states of light nuclei with well-developed $\alpha$ cluster structures. However, $\alpha$ clusters are generally bound in the ground state of nuclei, particularly in neutron-rich nuclei with a neutron skin on the surface, and can not be emitted spontaneously. A new reaction probe for $\alpha$ clustering in nuclei that are stable against $\alpha$ decay is thus needed. The quasi-free $(p,p \alpha)$ reaction is the method of choice, which takes $\alpha$ clusters out of the mother nucleus with minimized effect from the residual. A schematic view of the $(p,p \alpha)$ reaction in normal kinematics is shown in Fig.~\ref{QFSppa}(left panel). A high-energy proton beam (several hundred MeV) is sent onto the target of interest. When the transferred energy to the preformed $\alpha$ cluster is large in comparison to the $\alpha$ separation energy, the $(p,p \alpha)$ reaction can be considered as quasi-free---namely, the $\alpha$ cluster is knocked out without being disturbed by the residual---and the reaction cross section thus provides a good measure of the strength of the preformed $\alpha$ cluster in the mother nucleus. It is worthwhile to mention that similar nucleon knockout reactions such as $(e,ep)$,$(p,2p)$,$(p,pn)$ have been well established probes for the single-particle structures of nuclei~\citep{WAKASA2017,AUMANN2021}. Actually, $(p,p \alpha)$ was widely used in the 1970s and 1980s to study the cluster structure in light stable nuclei \citep{ROOS1977,CAREY1981,CAREY1984}. During the past decade,important progresses have been achieved in both experimental techniques and theoretical tools, making the $(p,p \alpha)$ reaction a sensitive probe for nuclear clustering (e.g., Ref.~\citep{Lyu-PRC-2018,Yoshida-2019}).Particularly, the increasing availability of radioactive ion beams also opens up new opportunities to investigate clustering in the ground state of unstable nuclei and in neutron-rich nuclear matter in a general context~\citep{Freer-RMP-2018,KIMURA2016,Enyo-2018,oertel2017equations}.

\begin{figure}[!http]
\centering
\includegraphics[trim = 0 0 0 0,clip,angle=0,width=6.0in]{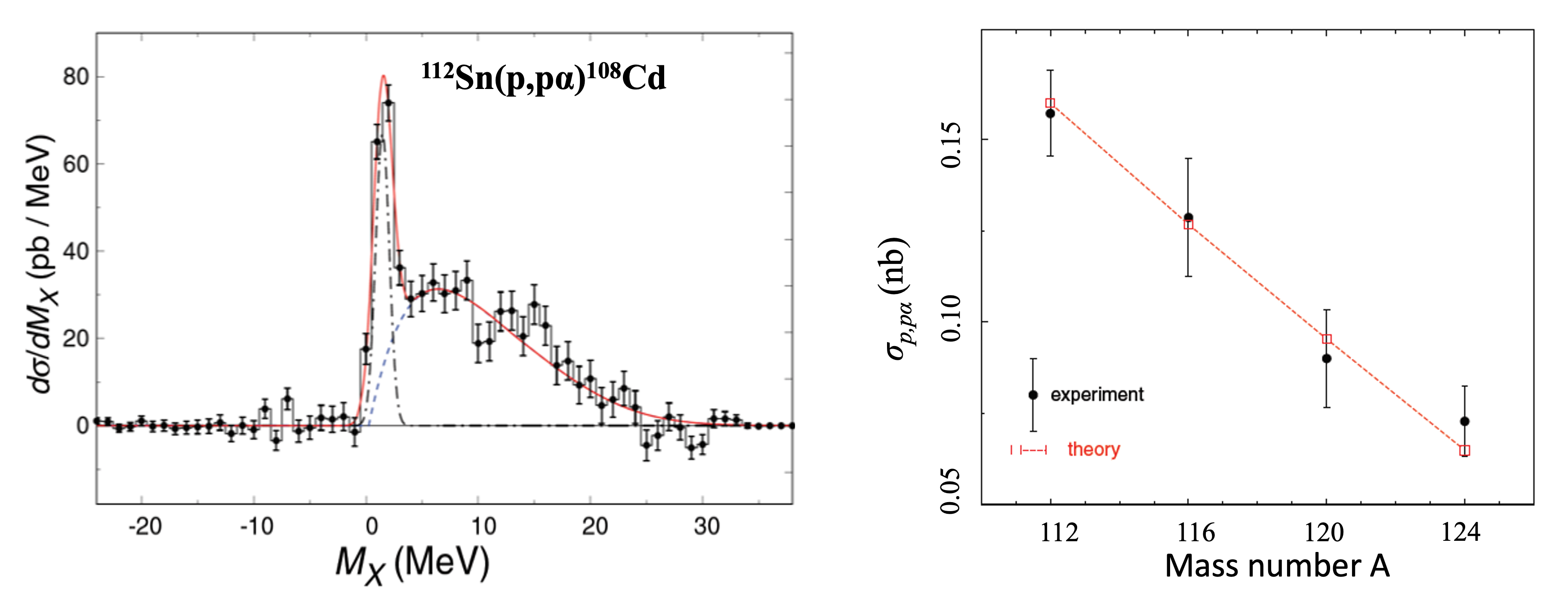}
\caption{\footnotesize{
The $\alpha$-separation-energy spectrum for the $\alpha$-knockout reaction $^{112}$Sn$(p,p \alpha)$ (upper panel), and the comparison of the experimental $\sigma_{p,p \alpha}$ cross section to the theoretical prediction (right panel). The figure is adapted from~\cite{TANAKA2021}.}}
\label{Snresults}
\end{figure}

In 2021, Tanaka {\it{et al.}} performed a quasi-free $(p,p \alpha)$ experiment at the Research Center for Nuclear Physics (RCNP) of Osaka University to measure the $\alpha$-clustering strength in stable tin isotopes $^{112,116,120,124}$Sn and its isotopic dependence~\citep{TANAKA2021}. For all the four tin isotopes, the $\alpha$-separation-energy spectrum [Fig.~\ref{Snresults} (left panel)] shows a clear peak located at the known $\alpha$-separation energy which is simply determined by the mass of the targe nucleus, as expected for the quasi-free knockout of preformed $\alpha$ clusters. This result thus provides direct evidence for the existence of preformed $\alpha$ clusters in these tin isotopes. As shown in the right panel of Fig.~\ref{Snresults}, $\sigma_{p,p \alpha}$ gradually decreases as the mass number increases. The observed isotopic systematics of $\sigma_{p,p \alpha}$ is well reproduced by reaction model calculations taking into account the radial density distributions of the $\alpha$ clusters of the gRDF prediction. This result reveals a tight interplay between the surface $\alpha$-clustering and the neutron skin in heavy nuclei and underlines the necessity of considering the effect of nuclear clustering when constraining the EOS parameters from the neutron-skin thickness~\citep{OERTEL2017}.

\begin{figure}[!http]
\centering
\includegraphics[trim = 0 0 0 0,clip,angle=0,width=4.0in]{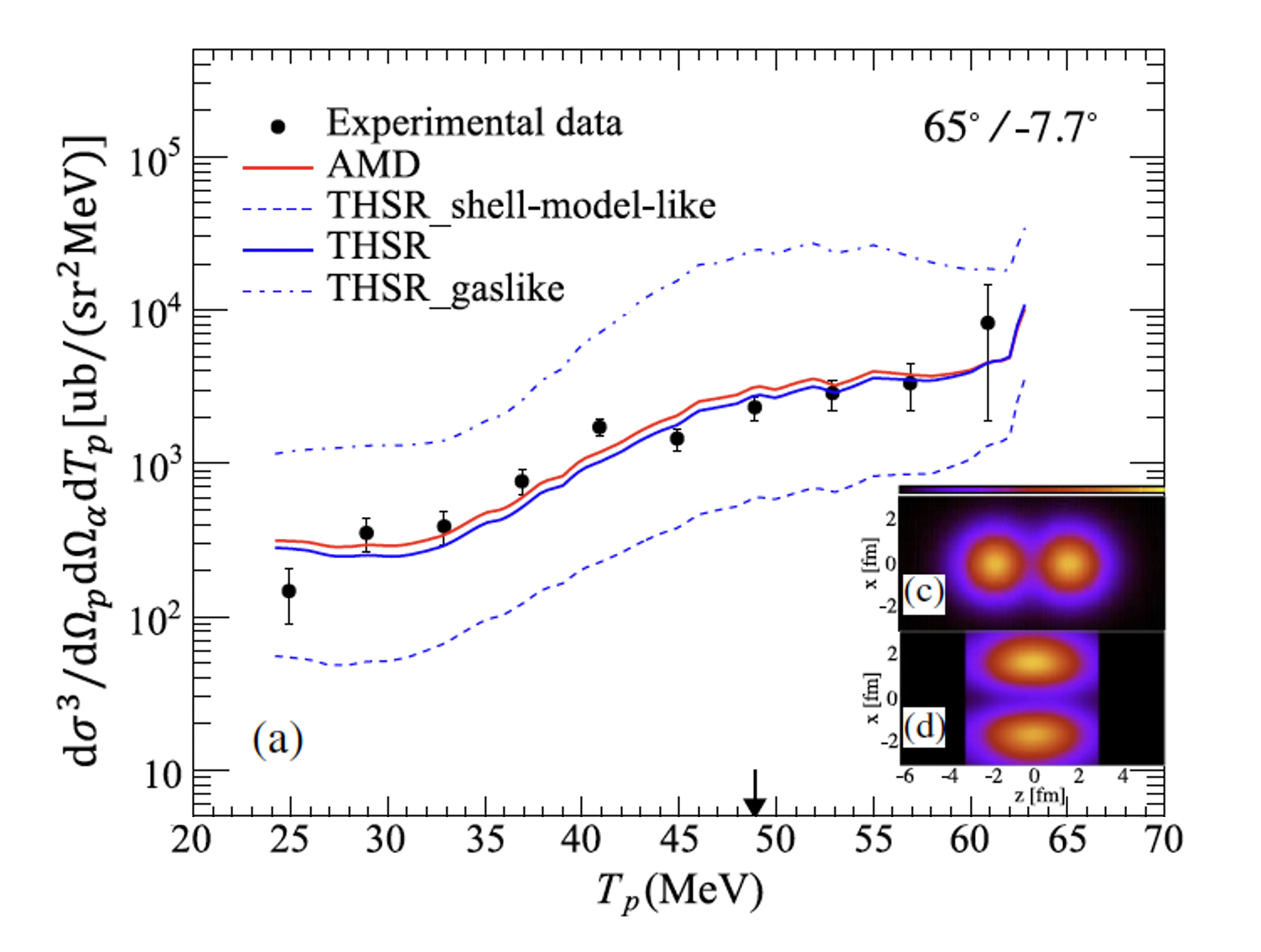}
\caption{\footnotesize{The triple differential cross section for $^{10}$Be $(p,p \alpha)$ $^{6}$He(g.s.), in comparison to theoretical calculations. The angles of the proton ($\ensuremath{\theta}_p$) and alpha ($\ensuremath{\theta}_{\alpha}$) are chosen according to the recoilless condition of the residual. The inset shows the density distribution of the protons and valence neutrons predicted by THSR, exhibiting a well-developed $\alpha$-$2n$-$\alpha$ cluster structure. The figure is adapted from~\citep{Li-ppa-2023}.} }
\label{QFSresult2}
\end{figure}

Recently, Li {\it{et al.}} performed the $(p,p \alpha)$ reaction in inverse kinematics (schematic view in Fig.~\ref{QFSppa}(right panel)) on the neutron-rich unstable nucleus $^{10}$Be~\cite{Li-ppa-2023}. According to the prediciton of microscopic cluster models THSR~\citep{Lyu-PRC-2018} and AMD~\citep{Enyo2015prc}, the ground state of $^{10}$Be has a well-developed $\alpha$-$2n$-$\alpha$ cluster structure---a dumbbell-like configuration of two $\alpha$ cores surrounded by two valence neutrons occupying the $\pi$ orbit. The $^{8}\mathrm{He}(p,p \alpha)$ experiment of Li {\it{et al.}} was performed at RIBF, and the experimental setup is based on the SAMURAI spectrometer. A 2-mm-thick solid hydrogen target (SHT) was used to enhance the luminosity while keeping the resolution at a level of $\sim1$ MeV in sigma. The recoil proton and the knocked-out $\alpha$ particle were measured in coincidence. From the measured angle and energy of the recoil proton and $\alpha$ particle, the $\alpha$-separation energy can be extracted by using the missing-mass method and different cluster configurations (leading to different states of the residual $^{6}$He) can thus be identified. In the experiment of Li {\it{et al.}}, the two reaction channels---$^{10}$Be$(p,p \alpha)$$^{6}$He(g.s.), $^{10}$Be$(p,p \alpha)$$^{6}$He(e.x.)---can also be distinguished from the particle identification (PID) of the fragments detected by the SAMURAI spectrometer since all the excited states of $^{6}$He are unbound. As shown in Fig.~\ref{QFSresult2}, the experimental triple differential cross section for the $^{10}$Be$(p,p \alpha)$$^{6}$He(g.s.) channel is nicely reproduced by the DWIA calculations incorporating the microscopic $\alpha$-cluster wave function of THSR and AMD. The results of Li {\it{et al.}} provide strong evidence for the $\alpha$-$2n$-$\alpha$ molecule-like cluster structure of $^{10}$Be(g.s.).

\section{Summary and Prospect}
Remarkable progress has been made on neutron correlations and neutron clusters in the first twenty years of the new century, but more questions still remain to be answered. The existence of a tetraneutron resonance was supported by two missing-mass experiments \citep{kisamori2016candidate,duer2022observation}. From the experimental point of view, new experiments using different production and measurement methods---particularly an invariant-mass measurement with the decay neutrons directly detected---are needed. From the theoretical point of view, the apparent discrepancies between many state-of-the-art models should be resolved. In this respect, it is worthwhile to mention that, despite the conflicting results about tetraneutron resonance, many theoretical models consistently find that the characteristics of the four-neutron system are insensitive to the three-body force \citep{higgins2020nonresonant, li2019ab, lazauskas2022low}, but its essential role was stressed in the QMC and GFMC calculations \citep{pieper2003can,gandolfi2017trineutron}. A trineutron resonance was predicted by the $\textit{ab initio}$ NCGSM and GFMC calculations \citep{li2019ab,gandolfi2017trineutron}---both predicting a $^3n$ resonance even lower than $^4n$, hinting at the working interactions or correlations beyond two neutrons, but was not supported by the recent experiment of Miki $et$ $al.$ \citep{MikiPRL2024}. Future experiments should go beyond the energy and width of the multi-neutron system and measure the internal neutron correlations. Such many-body interactions or correlations could be enhanced in a system with more neutrons and may thus give rise to more pronounced resonant structures in heavier neutron clusters ($^{6}n$ and $^{8}n$). It is also interesting to consider a multi-neutron cluster in a nuclear environment such as the low-density surface of neutron-rich nuclei. The neutron correlations are expected to be enhanced under such conditions as observed for $^2n$\citep{matsuo2006spatial,kubota2020surface}, and the bosonic dineutron clusters could thus be formed which may further form a condensate-like cluster state with multiple dineutron clusters---for example, the recently observed $\alpha+^2n+^2n$ condensate-like structure of the $0_2^{+}$ state of $^{8}\mathrm{He}$\citep{Yang2023prl}. Experimental measurements of the four-neutron or even six-neutron decay in extremely neutron-rich nuclei (e.g., $^{7}\mathrm{H}$ and $^{10}\mathrm{He}$) are also in progress at RIBF \citep{huang2021experimental}. It is worthwhile to mention that the multi-neutron detection efficiency decreases markedly as the number of neutrons increases. Neutron detector arrays with better resolution and higher efficiency for detection of multiple neutrons is essential. Moreover, an advanced multi-neutron identification algorithm is indispensable for the correct identification of true neutron signals out of the crosstalk signals \citep{huang2021experimental,KONDO2020}. 

Formation of $\alpha$ clusters in the low-density nuclear environment (significantly lower than the saturation density)---such as the surface region of heavy nuclei---has been well established in recent years (e.g., \citep{Hag12,Typel-2010,TYPEL2014,TANAKA2021}). Formation of clusters could also be favored in the ground state of light neutron-rich nuclei such as beryllium and carbon isotopes which can further give rise to molecular cluster structures, as predicted by microscopic cluster model calculations \cite{Freer-RMP-2018,KIMURA2016,Enyo-2018} and recently observed in $^{10}$Be \citep{Li-ppa-2023}. It is interesting to probe these novel structure phenomena using the quasi-free $(p,p \alpha)$ reaction. In addition, neutron-rich nuclei with a mixed state of clusters and neutrons could also serve as a unique laboratory for investigating the properties of neutron-rich matter, particularly when considering the inhomogeneity due to the formation of clusters. $\alpha$ clustering in heavy nuclei is essentially connected to $\alpha$ decay. While the cluster structure in light nuclei can be nicely described by microscopic cluster models, such as AMD \citep{KIMURA2016,Enyo-2018}and THSR \citep{Zhou2023nc}, and $ab$ initio nuclear models \citep{Otsuka2022nc,Elhatisari17}, it is still a challenge to quantitatively predict the $\alpha$ clustering strength in medium and heavy nuclei. Various theoretical approaches have been developed to predict the half-life of $\alpha$ decay and, more importantly, the preformation of $\alpha$ clusters (e.g., \citep{DELION1992,Betan2012,DONG2021,QIAN2018,XuChang2016}). The recent reanalysis of the $^{48}$Ti($p$, $p\alpha$)$^{44}$Ca data shows that AMD gives a two orders of magnitude smaller strength for $^{48}$Ti \citep{Yasutaka-2021}. This could be due to the strong spin-orbit interaction or the neglect of the tensor interaction in the AMD calculation, as discussed in \citep{uesaka2024}. It has been found in shell model calculations that the $nn$ and $pp$ pairing play a critical role in the formation of $\alpha$ clusters, but it is not clear about the role of the $np$ correlation \citep{QI2019}. In the future $(p,p \alpha)$ cluster knockout experiments should be applied to more neutron-rich nuclei and $\alpha$-radioactive heavy nuclei to probe the formation of $\alpha$ clusters in a broader context (see also discussions in \citep{uesaka2024,YANG2021SB}). It would also be interesting to investigate the formation of other light clusters like the deuteron, triton, and $^3$He by using similar quasi-free knockout reactions, which are predicted to behave similarly to the $\alpha$ cluster in both finite nuclei and nuclear matter~\citep{Typel-2010,LWChen-2017,OERTEL2017}. To comprehensively investigate clustering in medium-to-heavy-mass stable and unstable nuclei using cluster knockout reactions, a new research project ONOKORO led by Uesaka $et$ $al.$ has been launched at RIBF \citep{uesaka2024}. A dedicated detector system TOGAXSI is now under development for $(p,pX)$ cluster knockout reactions in inverse kinematics on unstable nuclei~\citep{TANAKA2023}. 

The operating and forthcoming RIB facilities worldwide, such as RIBF (Japan), FRIB (USA), HIAF (China), FAIR (Germany), and RAON (Korea), will provide great opportunities to study the cluster structure and neutron correlations of neutron-rich nuclei, and multi-neutron clusters. The concerted effort of experiment and theory would eventually elucidate the nature of nuclear clusters and the mechanism of their formation in neutron-rich nuclear systems.

\section*{Acknowledgment}

Z.Y. acknowledges the support from the National Key R$\&$D Program of China (Grants No. 2023YFE0101500 and 2022YFA1605100), the National Natural Science Foundation of China (Grant No. 12275006), and the State Key Laboratory of Nuclear Physics and Technology, Peking University (Grants No. NPT2022ZZ02 and NPT2024ZX01).


\bibliographystyle{ptephy}
\bibliography{neutron}
%

\vspace{0.2cm}
\noindent


\let\doi\relax








\end{document}